\documentstyle[12pt,twoside]{article}
{\catcode `\@=11 \global\let\AddToReset=\@addtoreset}
\AddToReset{equation}{section}
\renewcommand{\theequation}{\thesection.\arabic{equation}} 
 
\newcommand{\sgn}{{\rm sgn}}
\def\greaterthansquiggle{\raise.3ex\hbox{$>$\kern-.75em\lower1ex\hbox{$\sim$}}}
\def\lessthansquiggle{\raise.3ex\hbox{$<$\kern-.75em\lower1ex\hbox{$\sim$}}}
\newcommand{\beq}{\begin{equation}}
\newcommand{\eeq}{\end{equation}}
\newcommand{\beqa}{\begin{eqnarray}}
\newcommand{\eeqa}{\end{eqnarray}}
\newcommand{\beqan}{\begin{eqnarray*}}
\newcommand{\eeqan}{\end{eqnarray*}}
\newcommand{\ba}{\begin{array}}
\newcommand{\ea}{\end{array}}
\newcommand{\no}{\nonumber}

\newcommand{\ol}{\overline}
\newcommand{\ra}{\rightarrow}

\newcommand{\dar}{\downarrow}
\newcommand{\ve}{\varepsilon}
\newcommand{\vp}{\varphi}

\newcommand{\wt}{\widetilde}

\newcommand{\A}{{\cal A}}

\newcommand{\C}{{\cal C}}

\newcommand{\F}{{\cal F}}

\newcommand{\Ha}{{\cal H}}

\newcommand{\T}{{\cal T}}

\newcommand{\V}{{\cal V}}

\newcommand{\st}{\stackrel}

\newcommand{\dsum}{\displaystyle \sum}
\newcommand{\dprod}{\displaystyle \prod}

\def\nz{\ifmmode {I\hskip -3pt N} \else {\hbox {$I\hskip -3pt N$}}\fi}
\def\zz{\ifmmode {Z\hskip -4.8pt Z} \else
       {\hbox {$Z\hskip -4.8pt Z$}}\fi}
\def\qz{\ifmmode {Q\hskip -5.0pt\vrule height6.0pt depth 0pt
       \hskip 6pt} \else {\hbox
       {$Q\hskip -5.0pt\vrule height6.0pt depth 0pt\hskip 6pt$}}\fi}
\def\rz{\ifmmode {I\hskip -3pt R} \else {\hbox {$I\hskip -3pt R$}}\fi}
\def\cz{\ifmmode {C\hskip -4.8pt\vrule height5.8pt\hskip 6.3pt} \else
       {\hbox {$C\hskip -4.8pt\vrule height5.8pt\hskip 6.3pt$}}\fi}
\normalsize

\normalsize
\begin{document}
\bibliographystyle{plain}
\begin{titlepage}
\begin{flushright}
UWThPh--1998--7\\
ESI--553--1998\\
May 12, 1998
\end{flushright}
\vspace*{3cm}
\begin{center}
{\Large \bf  Do Anyons Solve Heisenberg's Urgleichung \\ [10pt]
in One Dimension$^\star$}\\[50pt]
N. Ilieva$^{\ast,\sharp}$ and W. Thirring  \\
\medskip
Institut f\"ur Theoretische Physik \\ Universit\"at Wien\\
\smallskip
and \\
\smallskip
Erwin Schr\"odinger International Institute\\
for Mathematical Physics\\
\vfill
{\bf Abstract} \\
\end{center}
\vspace{0.4cm}
We construct solutions to the chiral Thirring model in the framework of
algebraic quantum field theory. We find that for all positive temperatures
there are fermionic solutions only if the coupling constant is $\lambda =
\sqrt{2(2n+1)\pi}, \, n\in \bf N$.
\vfill
{\footnotesize

$^\star$ Work supported in part by ``Fonds zur F\"orderung der
wissenschaftlichen Forschung in \"Osterreich" under grant P11287--PHY; {\it to
be published in  Eur. Phys. J. C}

$^\ast$ Permanent address: Institute for Nuclear Research and Nuclear Energy,
Bulgarian Academy of Sciences, Boul.Tzarigradsko Chaussee 72, 1784 Sofia,
Bulgaria

$^\sharp$ E--mail address: ilieva@pap.univie.ac.at}

\end{titlepage}

\begin{quotation}
\section*{{\it Dedication}}
{\it
F. Schwabl is well--known for his contributions in condensed matter physics and
his book on quantum mechanics. However he was also among the pioneers for 
solving $(1\!+\!1)$--dimensional quantum field theories and it is with pleasure 
that we dedicate this note to his $60^{th}$ birthday.}
\end{quotation}

\section{Introduction}
It is usually taken for granted that fermions should enter the basic formalism
of the fundamental theory of elementary particles, the ultimate version of this
opinion being Heisenbergs ``Urgleichung" \cite{Hei}, in which no bose fields are
present at all. The opposite point of view, namely that theory including only
observable fields, necessarily uncharged bosons, is capable of describing
evolution and symmetries of a physical system, is the kernel of algebraic
approach to QFT, due to Haag and Kastler \cite{H}. Actually, the question which
is thus posed and which is of principal importance is whether and in which
cases definite conclusions about the time evolution and symmetries of charged
fields can be drawn from the knowledge about the observables that is gained
through experiment. Furthermore, before claiming that an ``Urgleichung" of the
type
\beq
\not\!\partial\psi(x) = \lambda\psi(x)\bar\psi(x)\psi(x)
\eeq
determines the whole Universe one should see whether it determines anything
mathematically.

Two--dimensional models offer a possibility to get a better feeling for these
problems due to the bose--fermi duality which takes place in two--dimensional
spacetime. This phenomenon amounts to the fact that in certain models formal
functions of fermi fields can be written that have vacuum expectation values
and statistics of bosons and vice versa. The equivalence is understood within
perturbation theory: the perturbation series for the so--related theories are
term--by--term equivalent (they may perfectly well exist even if the models are
not exactly solvable or if their physical sensibility is doubtful).

There are two facts which make such a duality possible. First comes the main
reason why soluble fermion models exist in two--dimensions, that is that
fermion currents can be constructed as ``fields" acting on the representation
space for the fermions. Also, the ``bosons into fermions" programme rests on
the fact that bosons in question are just the currents and fermions are
essentially determined by their commutation relations with them. Second comes
the observation which has been made in the pioneering works by Jordan \cite{J}
and Born \cite{BNN}: due to the unboundedness from below of the free--fermion
Hamiltonian the fermion creation and annihilation operators must undergo what
we should call now a Bogoliubov transformation which in addition leads to the
appearance of an anomalous term (later called ``Schwinger term") in the current
commutator, that in turn actually enables the ``bosonization". 

The ``fermions into bosons" part of the bose--fermi duality is fairly well
established, so that consistent expressions exist for the fermion bilinears
that are directly related to the observables of the theory.

The problem of rigorous definitions of operator--valued distributions and
eventually operators having the basic properties of fermion fields by taking
functions of bosonic fields is rather more delicate. On the level of operator
valued distributions solutions have been given by Dell'Antonio et al.\cite{DA}
and Mandelstam \cite{M} and on the level of operators in a Hilbert space --- by
Carey and collaborators \cite{CR, CHB} and in a Krein space by Acerbi, Morchio
and Strocchi \cite{AMS}.

Our goal is to see what elements are needed to make a solution of an equation
of the type (1.1) well defined. We shall not only reduce it to (1+1) dimensions
but will consider only one chiral component (a left or right mover) $\psi(x)$,
where $ \,x\,$ stands for $\,t\pm x\,$. Thus the question is how one can give a
precise meaning to the three ingredients
\beq
\begin{array}{llcl}
(a) & [\,\psi^\ast(x), \psi(x')\,]_+ = \delta(x\!-\!x'), \qquad 
[\,\psi(x), \psi(x')\,]_+ = 0 & \qquad & {\rm CAR} \\[7pt]
(b) & \frac{1}{i}\frac{d}{dx}\psi(x) = \lambda j(x)\psi(x) & 
\qquad & {\rm Urgleichung}\\[7pt]
(c) & j(x) = \psi^\ast(x)\psi(x) & \qquad & {\rm Current}
\end{array}
\eeq

Since (1.2b) involves derivatives of objects which are according to (1.2a)
rather discontinuous it is expedient to pass right away to the level of
operators in Hilbert space since there are plenty of topologies to control the
limiting procedures. In general norm convergence can hardly be hoped for but we
have to strive at least for strong convergence such that the limit of the
product is the product of the limits. With $\psi_f = \int_{-\infty}^{\infty}dx
f(x)\psi(x)$, (1.2a) becomes
\beq
[\,\psi^\ast_f, \psi_g\,]_+ = \langle f\vert g\rangle
\eeq
for $f\in L^2(\bf R)$ and $\langle .\vert .\rangle$ the scalar product in
$L^2(\bf R)$. This shows that $\psi_f$'s are bounded and form the
$C^\ast$--algebra CAR. There the translations $\,x\ra x+t\,$ give an automorphism
$\tau_t$ and we shall use the corresponding KMS--states $\omega_\beta$ and the
associated representation $\pi_\beta$ to extend CAR. Though there $j = \infty$,
one can give a meaning to $j$ as a strong limit in $\Ha_\beta$ by smearing
$\psi(x)$ over a region $\ve$ to $\psi_\ve(x)$ and define
$$
j_f = \int dx f(x) \lim_{\ve\to 0}\left(\psi^*_\ve(x)\psi_\ve(x) -
\omega_\beta(\psi^*_\ve(x)\psi_\ve(x))\right), \qquad f:{\bf R}\to {\bf R}
$$
These limits exist in the strong resolvent sense and define self--adjoint
operators which determine with
\beq
e^{ij_f}e^{ij_g} = e^{\frac{i}{8\pi}\int dx(f(x)g'(x)-f'(x)g(x))}e^{ij_{f+g}}
\eeq
the current algebra $\A_c$. Its Weyl structure is the same for all $\beta > 0$ and
$\omega_\beta$ extends to $\A_c$.

To construct the interacting fermions which on the level of distributions look
like 
$$
\Psi(x) = Z \,e^{i\lambda\int_{-\infty}^{x}dx'j(x')}
$$ 
(with some renormalization constant $Z$) poses two problems, one infrared and one
ultraviolet. For
\pagebreak
$$
\Psi_{\ve, R}(x) = e^{i\lambda\int dx'(\vp_\ve (x-x')-\vp_\ve (x-x'+R))
j(x')}, \qquad 
\vp_\ve(x) := \left\{\ba{cl} 1 & \mbox{ for } x \leq -\ve \\
-x/\ve & \mbox{ for } - \ve \leq x \leq 0 \\
0 & \mbox{ for } x \geq 0 \ea \right.
$$
neither the limit $R\to\infty$ nor the limit $\ve\to 0$ exist even as weak 
limits in $\Ha_\beta$. Thus one has to extend $\pi(\A_c)''$ to accomodate 
this kind of objects.

There are two equivalent ways of handling the infrared problem. Since the
automorphism generated by the unitaries $\Psi_{\ve, R}(x)$ converges to a
limit $\gamma$ for $R\to\infty$, one can form with it the crossed product 
$ \bar\A_c = \A_c \,\st{\gamma}{\bowtie}\,\rm Z$, so that in $\bar\A_c$ there
are unitaries with the properties which the limit should have. On the other
hand, the symplectic form in (1.4) and the state $\omega_\beta$ can be defined
for the limiting element $\Psi_\ve(x)$. This is what we will do in the text but
we also follow the former route in Appendix
B. In  any case $\bar\Ha_\beta$ assumes a sectorial structure, the subspaces  
$\A_c\,\dprod_{i=1}^{n} \Psi_\ve(x_i)\vert\Omega\rangle\,$ for different
$n$ are orthogonal and thus may be called $n$--fold charged sectors. The
$\Psi_\ve(x)$'s have the property that for $\vert x_i - x_j\vert > 2\ve$
they obey anyon statistics with parameter $\lambda^2$ and an Urgleichung (1.2b)
where $j(x)$ is averaged over a region of lenght $\ve$ below $x$. 

Removing the ultraviolet cut--off , $\ve\dar 0$, one could proceed as
before but in this case the sectors abound and the subspaces
$\,\A_c\Psi(x)\vert\Omega\rangle\,$ become orthogonal for different $x$, so
$\bar\Ha_\beta$ becomes non--separable. To get canonical fields of the type
(1.3) one has to combine $\ve\dar 0$ with a field renormalization
$\Psi_\ve \to \ve^{-1/2}\Psi_\ve$ such that
$$ 
\lim_{\ve\dar 0}\ve^{-1/2}\int dx f(x)\Psi_\ve(x) = \Psi_f 
$$
converge strongly in $\bar\Ha_\beta$ and satisfy (1.2b) in sense of
distributions.
However, they are not fermions but anyons and only for $\lambda = 
\sqrt{2(2n+1)\pi},\, n\in\bf N$ they are fermions. Thus we find that there is
indeed some magic about the Urgleichung inasmuch as on the quantum level it
allows fermionic solutions by this construction only for isolated values of the coupling
constant $\lambda$ whereas classically $\Psi(x) = Z \,e^{i\lambda\int_
{-\infty}^{x}dx'j(x')}$ solves (1.2b) for any $\lambda$. This feature can certainly not be seen by any power
expansion in $\lambda$.

The current (1.2c) has been constructed with the bare fermions $\psi$ and since
(1.2c) is sensitive to the infinite renormalization in the dressed field $\Psi$
it is better to replace (1.2c) by the requirement that $j_f$ is the generator
of a local gauge transformation. Indeed, 
\beq
e^{ij_f}\Psi_g e^{-ij_f} = \Psi_{e^{if}g}
\eeq
holds and in this sense (1.2c) is also satisfied.   
\vspace{06.cm}

\section{The CAR-algebra, its KMS-states and associated v. Neumann
algebras}
We start with the operator-valued distributions $\psi(x)$, $x \in {\bf R}$
which satisfy
\beq
[\psi^*(x),\psi(x')]_+ = \delta(x\!-\!x').
\eeq
For $f \in L^2({\bf R})$ we define the bounded operators
\beq
\psi_f = \int_{-\infty}^\infty dx \psi(x) f(x) =
\int_{-\infty}^\infty \frac{dp}{2\pi} \wt \psi(p) \wt f(p), \qquad
\wt f(p) = \int_{-\infty}^\infty dx \; e^{ipx} f(x)
\eeq
which form a C*-algebra $\A$ characterized by
\beq
[\psi^*_f,\psi_g]_+ = \langle f|g\rangle = \int dx f^*(x) g(x).
\eeq
We are interested in the automorphisms translation $\tau_t$ and parity $P$ and the
antiautomorphism charge conjugation $C$:
\beq
\tau_t \psi_f = \psi_{f_t}, \quad
f_t(x) = f(x-t), \quad
P\psi_f = \psi_{Pf}, \quad
P f(x) = f(-x), \quad
C \psi_f = \psi^*_f.
\eeq
$\A\,$ inherits the norm from $L^2({\bf R})$ such that $\,\tau_t\,$ is
(pointwise) normcontinuous in $\,t\,$ and even normdifferentiable for the
dense set of $f$'s for which
$$
\lim_{\delta \dar 0} \frac{f(x+\delta)-f(x)}{\delta} = f'(x)
$$
exists in $L^2({\bf R})$
\beq
\left.\frac{d}{dt} \; \tau_t \psi_f \right|_{t=0} = - \psi_{f'}.
\eeq
The $\tau$-KMS-states over $\A$ are given by
\beq
\omega_\beta(\psi^*_f \psi_g) = 
\int_{-\infty}^\infty \frac{dp}{2\pi} \frac{\wt f^*(p) \wt g(p)}
{1 + e^{\beta p}} =
\sum_{n=-\infty}^\infty \frac{(-1)^n}{2\pi}
\int \frac{dx dx' f^*(x) g(x')}{i(x-x') - n\beta + \ve}, \qquad
\ve \dar 0,
\eeq
$$
\omega_\beta(\psi_g \psi^*_f) = 
\omega_\beta(\psi^*_f \tau_{i\beta} \psi_g).
$$
With each $\omega_\beta$ are associated a representation $\pi_\beta$ with
cyclic vector $|\Omega\rangle$, $\omega(a) = \langle \Omega|a|\Omega
\rangle$ in $\Ha_\beta = \ol{\A|\Omega\rangle}$ and a v.~Neumann algebra
$\pi_\beta(\A)''$. It contains the current algebra $\A_c$ which gives
the formal expression $j(x) = \psi^*(x) \psi(x)$ a precise meaning.
We first observe
\paragraph{Lemma} (2.7) \\
If the kernel $K(k,k'): {\bf R}^2 \ra C$ is as operator $\geq 0$ and
trace class $(K(k,k) \in L^1({\bf R}))$, then $\forall \; \beta \in
{\bf R}^+$
\beqan
\lim_{M \ra \pm \infty} B_M &:=& \lim_{M \ra \pm \infty}
\frac{1}{(2\pi)^2}\int dk dk' K(k,k') \wt \psi^*(k+M) \wt \psi(k'+M) = \\
&=& \frac{1}{(2\pi)^2}\int dk dk' \lim_{M \ra \pm \infty} K(k,k') \omega_\beta
(\wt \psi^*(k+M) \wt \psi(k'+M)) = \\
&=& \left\{\ba{cl} \frac{1}{2\pi}\int dk \; K(k,k) & \mbox{ for } M \ra +\infty \\
0 & \mbox{ for } M \ra - \infty \ea \right.
\eeqan
in the strong sense in $\Ha_\beta$.
\paragraph{Remarks} (2.8)
\begin{enumerate}
\item (2.7) substantiates the feeling that for $k > 0$ most levels are
empty and for $k < 0$ most are full.
\item $B_M$ is a positive operator and by diagonalizing $K$ one sees
$$
\|B_M\| = \|K\|_1 = \frac{1}{2\pi}\int dk\; K(k,k).
$$
\end{enumerate}
\paragraph{Proof:} Since the norms of $B_M$ are bounded uniformly for all
$M\,$, it is sufficient to show strong convergence on a dense set in $\Ha_\beta$.
Furthermore
$$
\|A_Ma|\Omega\rangle\|^2 = 
\langle \Omega|A^*_M A_M a \tau_{i\beta} a^*|\Omega\rangle 
\leq \|A_M\Omega\rangle \| \|A_M a \tau_{i\beta} a^*|\Omega\rangle\|
\qquad \forall a\in\A.
$$
Thus if $\|A_M|\Omega\rangle\| \ra 0$ and $\|A_M\|$ uniformly bounded,
then $A_M \ra 0$ since with $a \in \A\,$,  $\|a \tau_{i\beta}a^*|\Omega\rangle\| < \infty$
are dense in $\Ha_\beta$. Now
$$
\langle \Omega|(B_M - \langle B_M\rangle)^2|\Omega\rangle =
\langle \Omega| B^2_M|\Omega\rangle - \langle \Omega|B_M|\Omega\rangle^2
$$
contains the distributions
$$
\langle \Omega|\wt \psi(k+M)^* \wt \psi(k'+M) \wt \psi(q'+M)^* 
\wt \psi(q+M)|\Omega\rangle - \langle \Omega|\cdot|\Omega\rangle \;
\langle \Omega|\cdot|\Omega\rangle = 
$$
$$
= \frac{(2\pi)^2 \delta(k\!-\!q) \delta(k'\!-\!q')}{(1 + e^{\beta(k+M)})(1 + 
e^{-\beta(k'+M)}}.
$$
This gives for the operators
$$
\langle \Omega|B_M^2|\Omega\rangle - \langle \Omega|B_M|\Omega\rangle^2
= \frac{1}{(2\pi)^2}\int \frac{dk dk'|K(k,k')|^2}{(1 + e^{\beta(k+M)})
(1 + e^{-\beta(k'+M)})} \eqno(2.9).
$$
Since the Hilbert--Schmidt norm $\int K^2 < \infty$ is less than the trace norm 
and 
the integrand in (2.9) for $M \ra \pm \infty$  goes to zero uniformly on compact
sets we have established $B_M \ra \langle B_M\rangle$ for $M \ra \pm \infty$.

If $\int |K|^2$ keeps increasing with $M$, then $B_M - \langle B_M\rangle$ 
may nevertheless tend to an (unbounded) operator.

\paragraph{Lemma} (2.10) \\
If
$$
B_M = \frac{1}{(2\pi)^2}\int dk dk' \wt f(k-k') \Theta(M-|k|) 
\Theta(M-|k'|) \wt \psi^*(k) \psi(k')
$$
with $\wt f$ decreasing faster than an exponential and being the Fourier 
transform 
of a positive function, the $B_M - \omega_\beta(B_M)$ is a strong Cauchy
sequence for $M \ra \infty$ on a dense domain on $\Ha_\beta$.

\paragraph{Remarks} (2.11)
\begin{enumerate}
\item From (2.8) we know that $\|B_M\| < 2M \wt f(0)$ and $f(x) \geq 0$ is not 
a serious restriction since any function is a linear combination of positive
functions.
\item Since the limit $j_f$ is unbounded the convergence is not on all
of $\Ha_\beta$, however since for the limit $j_f$ holds $\tau_{i\beta}j_f = 
j_{e^{\beta p} f}$,  the dense domain is invariant under $j_f$. Thus we have
strong resolvent convergence which means that bounded functions of
$B_M$ converge strongly. Also the commutator of the limit is the limit
of the commutators.
\end{enumerate}
\paragraph{Proof:} As before
\beqan
\lefteqn{\langle \Omega|(B_{M'} - B_M - \omega(B_{M'} - B_M))^2
|\Omega\rangle = } \\
&=& \int_{-\infty}^\infty \frac{dk dk'}{(2\pi)^2} |\wt f(k-k')|^2
(1 + e^{\beta k})^{-1} (1 + e^{-\beta k'})^{-1} \cdot \\
&& \mbox{} \cdot \left[\Theta(M'-|k|) \Theta(M'-|k'|) - \Theta(M-|k|)
\Theta(M'-|k'|)\right]
\eeqan
for $M' > M$. Now with $q = k'-k$ we have
$$
\int_M^{M'} \frac{dk}{(1 + e^{\beta k})(1 + e^{-\beta(k+q)})} 
\leq \int_M^{M'} dk \; e^{-\beta k}
$$
and similarly for $\int_{-M'}^{-M} dk$. Altogether we get
$$
\leq \int \frac{dq}{2\pi} |f(q)|^2 \frac{1 + e^{\beta|q|}}{2}
(e^{-\beta M} - e^{-\beta M'}).
$$ 
By assumption $\int dq < \infty$ thus
$\forall \; \ve > 0\,$ $\exists \; M$ such that this is $< \ve$ $\,\forall 
M' > M$.

We conclude that the limit exists and is selfadjoint on a suitable domain.
We shall write it formally
$$
j_f = \int_{-\infty}^\infty \frac{dk dk'}{(2\pi)^2} \; \wt f(k-k')
: \wt \psi(k)^* \wt \psi(k'):  \eqno(2.12)
$$

Next we show that the currents so defined satisfy the CCR with a suitable symplectic form
$\sigma$ \cite{J, Schwinger}.
\paragraph{Theorem} (2.13)
$$
[j_f,j_g] = i \sigma(f,g) =
\int_{-\infty}^\infty \frac{dp}{(2\pi)^2} \; p \wt f(p) \wt g(-p) =
\frac{i}{4\pi} \int_{-\infty}^\infty dx(f'(x)g(x) - f(x)g'(x)).
$$
\paragraph{Proof:} For the distributions $\wt \psi(k)$ we get algebraically
$$
[\wt\psi^*(k) \wt\psi(k'),\wt\psi^*(q)\wt\psi(q')] =
2\pi\left[\wt\psi^*(k)\wt\psi(q')\delta(q\!-\!k') -
\wt\psi^*(q)\wt\psi(k') \delta(k\!-\!q')\right]
$$
and for the operators after some change of variables
$$
\frac{1}{(2\pi)^3}\int dkdpdp' \wt f(p) \wt g(p') \wt\psi^*(p+p'+k) \wt\psi(k)
\Theta(M-|k|) \Theta(M-|p+p'+k|) \cdot
$$
$$
\cdot \left[\Theta(M-|p'+k|) - \Theta(M-|p+k|)\right].
$$
For fixed $p$ and $p'$ and $M \ra \infty$ we see that the allowed region
for $k$ is contained in $(M-|p|-|p'|,M)$ and
$(-M,-M+|p|+|p'|)$. Upon $k \ra k \pm M$ we are in the situation of (2.7),
thus we see that the commutator of the currents (2.12)
is bounded uniformly in $M$ if $\wt f$ and 
$\wt g$ decay faster than exponentials and converges to the expectation
value. This gives finally
$$
\int_{-\infty}^\infty \frac{dp}{(2\pi)^2} \; \wt f(p) \wt g(-p)
\int dk \; \Theta(M-|k|) \left[\Theta(M - |k-p|) - \Theta(M - |k+p|)\right]
\frac{1}{1 + e^{\beta k}}
$$
$$
\st{M \ra \infty}{\longrightarrow} \int_{-\infty}^\infty \frac{dp}{(2\pi)^2}
\; p \wt f(p) \wt g(-p).
$$

\paragraph{Remarks} (2.14)
\begin{enumerate}
\item Since the $j_f$'s satisfy the CCR they cannot be bounded and it is
better to write (2.12) in the Weyl form for the associated unitaries
$$
e^{ij_f} \; e^{ij_g} = e^{\frac{i}{2} \sigma(g,f)} \; e^{ij_{f+g}} =
e^{i\sigma(g,f)} \; e^{ij_g} \; e^{ij_f}.
$$ 
\item Since $j_f$ is selfadjoint, $\,e^{i\alpha j_f}$ generate 1--parameter
groups. They are the local gauge transformations
$$
e^{-i\alpha j_f} \; \psi_g \; e^{i\alpha j_f} = \psi_{e^{i\alpha f}g}.
$$
\item The state $\,\omega_\beta\,$ can be extended to $\,\bar \omega_\beta\,$
over $\,\pi_\beta(\A)''$ and $\tau_t$ to $\bar \tau_t,\,\,
\bar \tau_t \in \mbox{Aut }\pi_\beta(\A)''$ with $\bar \tau_t \,j_f = j_{f_t}$.
Furthermore $\,\bar \omega_\beta\,$ is $\,\bar \tau$--KMS and is calculated to 
be (Appendix A, see also \cite{BY})
$$
\bar \omega_\beta(e^{ij_f}) = \exp \left[ -\frac{1}{2} \int_{-\infty}^\infty
\frac{dp}{(2\pi)^2} \frac{p}{1 - e^{-\beta p}} |\wt f(p)|^2  \right].
$$
\item $\bar\omega_\beta$ is not invariant under the parity $P$ (2.4).
This symmetry is destroyed in $\pi_\beta$,
$$
[j(x),j(x')] = -\,\frac{i}{2\pi}\,\delta'(x\!-\!x')
$$
is not invariant under $j(x) \ra j(-x)$. 
Thus $P \notin \mbox{Aut }\pi_\beta(\A)''$.
\item The extended shift automorphism $\,\bar \tau_t\,$ is not only strongly 
continuous but for
suitable $f$'s also differentiable in $\,t\,$ (strongly on a dense set in 
$\Ha_\beta$)
$$
\frac{1}{i} \frac{d}{dt} \bar \tau_t e^{ij_f} = 
\left[ j_{f'_t} + \frac{1}{2} \sigma(f_t,f'_t)  \right] e^{ij_{f_t}} =
e^{ij_{f_t}} \left[j_{f'_t} - \frac{1}{2} \sigma(f_t,f'_t) \right] =
\frac{1}{2} \left[ j_{f'_t} \; e^{ij_{f_t}} + e^{ij_{f_t}} j_{f'_t} \right].
$$
\item The symplectic structure is formally independent on $\beta$, however for 
$\beta < 0$ it changes its sign,
$\sigma \to -\sigma$, and for $\beta = 0$ (the tracial state) it becomes zero.
\end{enumerate}
\vspace{0.6cm}

\section{Extensions of $\A_c$}
So far $\A_c$ was defined for $j_f$'s with $\, f \in C_0^\infty$, for
instance. The algebraic structure is determined by the symplectic form 
$\sigma(f,g)$ (2.13) which 
is actually well defined also for the 
Sobolev space, $\sigma(f,g) \ra \sigma(\bar f, \bar g), \, \bar f, \bar g \in
H_1, \, H_1 = \{f : f,f' \in L^2\}\,$. Also $\bar \omega_\beta$
can be extended to $H_1$, since $\,\bar \omega_\beta(e^{ij_{\bar f}}) > 0\,$
for $\,\bar f \in H_1$.
However the anticommuting operators we are looking for are of the form
$e^{ij_f}$, $f(x) = 2\pi \Theta(x_0\!-\!x)$ and though one can give
$\sigma(f,g)$ a meaning for such an $f$,  one has in $\omega_\beta$ a 
divergence for $p \ra 0$ and $p \ra \infty$
$$
\omega_\beta(e^{ij_\Theta}) = \exp \left[-\frac{1}{2} \int_{-\infty}^\infty
\frac{dp}{p(1 - e^{-\beta p})} \right] = 0
$$
and thus $\langle f \vert e^{ij_\Theta} \vert f \rangle = 0\,$, where $\vert f
\rangle = e^{ij_f}\vert\Omega\rangle$ are total in $\Ha_\beta$. Thus this
operator acts as zero in $\Ha_\beta$.
If one tries to approximate $\Theta$ by functions from $H_1$, the unitaries
converge weakly to zero.
\paragraph{Example} (3.1) \\
Denote
$$
\vp_\ve(x) := \left\{\ba{cl} 1 & \mbox{ for } x \leq -\ve \\
-x/\ve & \mbox{ for } - \ve \leq x \leq 0 \\
0 & \mbox{ for } x \geq 0 \ea \right., \qquad
\Phi_{\delta,\ve}(x) := \vp_\ve(x) - \vp_\ve(x + \delta) \in H_1, 
$$
$$
\lim_{\delta \ra \infty \atop \ve \ra 0} \Phi_{\delta,\ve}(x) =
\Theta(x).
$$
Then
$$
\wt \Phi_{\delta,\ve}(p) = \frac{1 - e^{ip\ve}}{\ve p^2} (1 - e^{ip\delta})
$$
and
$$
\|\Phi_{\delta,\ve}\|^2_\beta = \int_{-\infty}^\infty \frac{dp}{2\pi} 
\frac{p}{1 - e^{-\beta p}}
| \wt \Phi(p)|^2 =
16 \int_{-\infty}^\infty \frac{dp}{2\pi} \frac{p}{1 - e^{-\beta p}}
\frac{\sin^2 p \ve/2}{\ve^2 p^4} \sin^2 p \delta/2 , 
$$
$$
\|\Phi\|^2_\beta \geq c \int_0^{1/\delta} dp \; \delta^2 = c \delta
$$
for $\beta/\delta$, $\ve/\delta \ll 1$ and $c$ a constant. Thus for
$\delta \ra \infty\,$, $\,\|\Phi_\delta\|_\beta \ra \infty$. Also $\,
\|\Phi_\delta - f\|_\beta \ra \infty$ since 
$$
\|\Phi_\delta - f\|_\beta \geq \|\Phi_\delta\|_\beta - \|f\|_\beta \ra \infty
\qquad \forall \; \|f\|_\beta < \infty
$$
and thus
$$
|\langle \Omega|e^{-ij_f} \; e^{ij_{\Phi_\delta}}|\Omega\rangle| =
e^{-\frac{1}{2} \|\Phi_\delta - f\|_\beta^2} \ra 0.
$$
But $e^{ij_f}|\Omega\rangle$, $\|f\|_\beta < \infty$, is total in 
$\Ha_\beta$ and thus $e^{ij_{\Phi_\delta}}|\Omega\rangle$ and therefore
$e^{ij_{\Phi_\delta}}$ goes weakly to zero. However the automorphism
$$
e^{ij_f} \ra e^{-ij_{\Phi_\delta}} e^{ij_f} e^{ij_{\Phi_\delta}} =
e^{i \sigma(\Phi_\delta,f)} e^{ij_f}
$$
converges since
$$
\sigma(f,\Phi_\delta) = - \frac{1}{2\pi\ve} \left( \int_{-\ve}^0 -
\int_{-\ve - \delta}^{-\delta}\right) dx \; f(x) 
\st{\delta \ra \infty}{\longrightarrow}
- \frac{1}{2\pi\ve} \int_{-\ve}^0 dx \; f(x)
\st{\ve\ra 0}{\longrightarrow}
-\frac{1}{2\pi}\,f(0).
$$
This divergence of $\|\Phi_{\delta,\ve}\|$ is related to the well--known
infrared problem of the massless scalar field in $(1+1)$ dimensions and
various remedies have been proposed \cite{S}. We take it as a sign
that one should enlarge $\,\A_c\,$ to some $\,\bar \A_c\,$ and work in the 
Hilbert space $\bar \Ha$ generated by $\,\bar \A_c\,$ on the natural extension 
of the state. Thus we add to
$\A_c$ the idealized element $\,e^{i 2\pi j_{\vp_\ve}} = U_\pi\,$
and keep $\sigma$ and $\omega_\beta$ as before.
Equivalently we take the automorphism $\gamma$ generated by $U_\pi$ and
consider the crossed product $\,\bar \A_c = \A_c \st{\gamma}{\bowtie} 
\bf Z\,$. There is a natural extension $\bar \omega$ to $\bar \A_c$ and a 
natural isomorphism of $\bar \Ha$ and $\,\bar \A_c|\bar \Omega\rangle$.
Here $\bar \Ha$ is the countable orthogonal sum of sectors with $n$
particles created by $U_\pi$. Thus,
$$
\langle \Omega|e^{ij_f} U_\pi|\Omega\rangle = 0 \eqno(3.2)
$$
means that $U_\pi$ leads to the one-particle sector, 
in general
$$
\langle \Omega|U_\pi^{*n} \, e^{ij_f} U_\pi^m|\Omega\rangle =
\delta_{nm}\,\omega_\beta(\gamma^n\,e^{ij_f}).
$$ 
The quasifree automorphisms on $\,\A_c$ (e.g. $\tau_t$) 
can be naturally extended to $\,\bar \A_c\,\,$, $\tau_t \,U_\pi =
e^{i\pi j_{\vp_{\ve,t}}}\,$, $\,\vp_{\ve,t}(x) = \vp_\ve(x+t)\,$  and since
$\vp_\ve - \vp_{\ve,t} \in H_1$ $\,\forall \; t$, this does not lead out
of $\bar \A_c$. 

$U_\pi$ has some features of a fermionic field since
$$
\sigma(\vp_\ve,\tau_t \vp_\ve) = - \sigma(\vp_\ve,\tau_{-t} \vp_\ve) =
\frac{1}{4\pi}\left\{\ba{cl} 1 & \mbox{ for } t > \ve \\
\frac{2t}{\ve} - \frac{t^2}{\ve^2} & \mbox{ for } 0 \leq t \leq \ve \ea \right. . \eqno(3.3)
$$
More generally we could define $U_\alpha = e^{i \sqrt{2\pi\alpha} j_{\vp_\ve}}$
and get from (3.3) with 
$$
\sgn(t) = \Theta(x) - \Theta(-x) = \left\{
\begin{array}{cl} 1 & {\rm for} \qquad t > 0 \\
0 & {\rm for} \qquad t = 0 \\
-1 & {\rm for} \qquad t <  0.
\end{array} \right.
$$
\paragraph{Proposition} (3.4)
\beqan
U_\alpha \tau_t U_\alpha &=& \tau_t(U_\alpha) U_\alpha \,
e^{i\,\alpha\,\sgn(t)/2}, \\
U^*_\alpha \tau_t U_\alpha &=& \tau_t(U_\alpha) U^*_\alpha \,
e^{i\,\alpha\,\sgn(t)/2} \quad \forall \; |t| > \ve.
\eeqan

\paragraph{Remark} (3.5) \\
We note a striking difference between the general case of anyon statistics 
and the two particular cases --- Bose ($\alpha = 2\cdot 2n\pi$) or Fermi 
($\alpha = 2(2n+1)\pi$) 
statistics. Only in the latter two cases parity $P$ (2.4) is an
automorphism of the extended algebra generated through $U_\alpha$. Thus $P$
which was destroyed in $\A_c$ is now recovered for two subalgebras.
The particle sectors are orthogonal in any case
$$
\langle \Omega| U^*{}^n_\alpha \,e^{ij_f} U^m_\alpha|\Omega\rangle =
0 \quad \forall \; n \neq m, \; f \in H_1.
$$
Furthermore, sectors with different statistics are orthogonal
$\langle\Omega\vert U^*_\alpha U_{\alpha'}\vert\Omega\rangle = 0, \, \alpha
\not= \alpha'$, thus if we adjoin $U_\alpha, \forall \alpha\in{\bf R}$,
$\bar\Ha_\beta$ becomes nonseparable.
 
Next we want to get rid of the ultraviolet cut--off and let $\ve$ go to
zero. Proceeding the same way we can extend $\sigma$ and $\tau_t$ but 
keeping $\omega$ the sectors abound. The reason is that
$\vp_\ve \st{\ve \ra 0}{\longrightarrow} \Theta(x)$ and
$$
\|\Theta - \Theta_t\|^2 = \int_{-\infty}^\infty 
\frac{dp \; p}{1 - e^{-\beta p}} \frac{|1 - e^{itp}|^2}{p^2}
$$
is finite near $p = 0$ but diverges logarithmically for $p \ra \infty$.
This means that $e^{ij_f} e^{ij_\Theta}|\Omega\rangle$, $f \in H_1$
gives a sector where one of these particles (fermions, bosons or anyons)
is at the point $x = 0$ and is orthogonal to $e^{ij_f} e^{ij_{\Theta_t}}
| \Omega\rangle$ $\forall \; t \not= 0$. Thus the total Hilbert space is not
separable and the shift $\tau_t$ is not even weakly continuous. Thus
there is no chance to make sense of $\frac{d}{dt} \tau_t e^{ij_\Theta}$.
\vspace{0.6cm}

\section{Anyon fields in $\pi_{\bar \omega}(\bar \A_c)''$}
Next we shall use another ultraviolet limit to construct local fields
which obey some anyon statistics. Of course quantities like
$$
[\Psi^*(x),\Psi(x')]_\alpha := 
\Psi^*(x) \Psi(x') e^{i \frac{2\pi-\alpha}{4} \sgn(x'-x)} 
+ \Psi(x') \Psi^*(x) e^{-i \frac{2\pi-\alpha}{4} \sgn(x'-x)} =
\delta(x\!-\!x')
$$
will only be operator valued distributions and have to be smeared to give
operators. Furthermore in this limit the unitaries we used so far have to 
be renormalized so that $\delta(x\!-\!x')$ gets a factor 1 in front. A candidate
for $\Psi(x)$ will be ($\alpha \in (0, 4\pi)$)
$$
\Psi(x) := 
\lim_{\ve \ra 0} n(\ve) \exp \left[ i \sqrt{2\pi\alpha} \int_{-\infty}^\infty
dy \; \vp_\ve(x-y) j(y) \right] 
$$
with $\vp_\ve$ from (3.1) and $n(\ve)$ a suitably chosen normalization.
With the shorthand $\vp_{\ve,x}(y) = \vp_\ve(x-y)$ we can write
\beqan
\Psi^*_\ve(x) \Psi_\ve(x') &=& 
\exp \left\{ i \,2\pi\alpha \,\sigma(\vp_{\ve,x},\vp_{\ve,x'}) \right\}
\exp \left\{ i \sqrt{2\pi\alpha} \, j_{\vp_{\ve,x'} - \vp_{\ve,x}}\right\}, \\
\Psi_\ve(x') \Psi^*_\ve(x) &=&
\exp \left\{- i \,2\pi\alpha \,\sigma(\vp_{\ve,x},\vp_{\ve,x'}) \right\}
\exp \left\{ i \,\sqrt{2\pi\alpha} \,j_{\vp_{\ve,x'} - \vp_{\ve,x}}\right\}.
\eeqan
We had in (3.3)
$$
4\pi\sigma(\vp_{\ve,x},\vp_{\ve,x'}) = \sgn(x-x')
\left\{ \Theta(|x-x'| - \ve) + \Theta(\ve - |x-x'|)
\frac{(x - x')^2}{\ve^2} \right\} 
$$
$$
=: \sgn(x-x') D_\ve(x-x')
$$
and thus
$$
[\Psi^*_\ve(x),\Psi_\ve(x')]_\alpha = 2n(\ve)^2 \cos \left[\sgn(x-x')
\left( \frac{\pi}{2} - \frac{\alpha}{4}(1 - D_\ve(x-x'))\right)\right]
\exp \left[ i \alpha j_{\vp_{\ve,x'} - \vp_{\ve,x}}\right].
$$
Note that for $|x-x'| \geq \ve\,$ the argument of the $\cos$ becomes 
$ \pm \pi/2$, so the $\alpha$--commutator vanishes, in agreement with (3.4). 
To manufacture a $\delta$-function for $|x-x'|
\leq \ve$ we note that $cos(...) > 0$ and $\omega_\beta(e^{i\alpha j}) > 0$, so 
we have to choose $n(\ve)$ such that
$$
2 n^2(\ve)\ve \int_{-1}^1 d\delta \cos \left( \frac{\pi}{2} -
\frac{\alpha}{4}(1 - \delta^2) \right) \cdot \omega_\beta
\left(\exp \left[ i \alpha j_{\vp_{\ve,x-\ve\delta} - \vp_{\ve,x}}\right]\right)
= 1
$$
and to verify that for $\ve \dar 0$ $\,[\;]_\alpha$ converges 
strongly to a $c$-number. For the latter to be finite we have to smear $\Psi(x)$ with
$L^2$-functions $g$ and $h$:
$$
\int dx dx' g^*(x) h(x') [\Psi^*_{\ve}(x),\Psi_{\ve}(x')]_\alpha = 
\int dx dx' g^*(x) h(x') 2n(\ve)^2 \cos(\;)
\exp \left[ i \alpha j_{\vp_{\ve,x'} - \vp_{\ve,x}}\right].
$$
This converges strongly to $\langle g|h\rangle$ if for $\ve \dar 0$
$$
\left\langle\exp \left[- i \alpha j_{\vp_{\ve,x'} - \vp_{\ve,x}}\right]
\exp \left[ i \alpha j_{\vp_{\ve,y'} - \vp_{\ve,y}}\right]\right\rangle
- \left\langle\exp \left[- i \alpha j_{\vp_{\ve,x'} - \vp_{\ve,x}}\right]
\right\rangle \left\langle
\exp \left[ i \alpha j_{\vp_{\ve,y'} - \vp_{\ve,y}}\right]\right\rangle
\ra 0
$$
for almost all $x,x',y,y'$. Now
$$
\langle e^{-ij_a} e^{ij_b} \rangle = \langle e^{-ij_a}\rangle
\langle e^{ij_b}\rangle
\exp \left[\int_{-\infty}^\infty \frac{dp \;p}{1 - e^{\beta p}}
\wt a(-p) \wt b(p) \right].
$$
In our case this last factor is
\beqan
\lefteqn{ \int_{-\infty}^\infty \frac{dp \;p}{1 - e^{-\beta p}}
\frac{|1 - e^{ip\ve}|^2}{\ve^2 p^4} (e^{ipx} - e^{ipx'})
(e^{-ipy} - e^{-ipy'}) = } \\
&=& \int_{-\infty}^\infty \frac{dp\;2(1-\cos p)}{p^3(1-e^{-\beta p/\ve})}
(e^{ipx/\ve} - e^{ipx'/\ve})(e^{-ipy/\ve} - e^{-ipy'/\ve}).
\eeqan
For fixed $\beta \neq 0$ and almost all $x,x',y,y'$ this converges to zero
for $\ve \ra 0$ by Riemann-Lebesgue. In the same way one sees that
$\exp \left[i \alpha j_{\vp_{\ve,x} + \vp_{\ve,x'}} \right]$ converges
strongly to zero and that the $\Psi_{\ve,g}$ are a strong Cauchy sequence
for $\ve \ra 0$. To summarize we state
\paragraph{Theorem} (4.1) \\
$\Psi_{\ve,g}$ converges strongly for $\ve \ra 0$ to an operator
$\Psi_g$ which for $\alpha = 2\pi$ satisfies 
$$
[\Psi_g^*,\Psi_h]_+ = \langle g|h\rangle, \qquad
[\Psi_g,\Psi_h]_+ = 0.
$$
If $supp\, g < supp\, h$, 
$$
\Psi_g^* \Psi_h \, e^{i\frac{2\pi - \alpha}{4}} +
\Psi_h \Psi_g^* \, e^{-i\frac{2\pi - \alpha}{4}} = 0 \qquad \forall \alpha.
$$

Furthermore we have to verify the claim (1.5) that also for $\Psi_g$ the
current $j_f$ induces the local gauge transformation $g(x) \ra e^{2i\alpha f(x)}
g(x)$. For finite $\ve$ we have 
$$
e^{ij_f} \Psi_{\ve,g} e^{-ij_f} = 
\Psi_{\ve, e^{i 2\pi\alpha \sigma(f,\vp_\ve)}g}
$$
and for $\ve \dar 0$ we get
$\sigma(f, \vp_\ve) \ra \frac{1}{2\pi}f(0)\,$, so that $\,\sigma(f,
\tau_x\vp_\ve) = \frac{1}{2\pi}f(x)$. \\[2pt]

To conclude we investigate the status of the ``Urgleichung'' in our
construction. It is clear that the product of operator valued distributions on
the r.h.s. can assume a meaning only by a definite limiting prescription.
Formally it would be
$$
\Psi(x)\Psi^*(x)\Psi(x) = [\Psi(x), \Psi^*(x)]_+\Psi(x) - 
\Psi^*(x)\Psi(x)^2 = \delta(0)\Psi(x) - 0.
$$
From (2.13,5) we know
$$
\frac{1}{i} \frac{\partial}{\partial x} \Psi_\ve(x) = 
\frac{\sqrt{2\pi\alpha}}{2} \,[\bar \jmath(x),\Psi_\ve(x)]_+, \qquad
\bar \jmath(x) = \frac{1}{\ve} \int_{x-\ve}^x dy \; j(y).
$$
Using $j_{\vp'} e^{ij_\phi} = \frac{1}{i}\frac{\partial}{\partial\alpha}
e^{i\frac{\alpha}{2}\sigma(\vp',\vp)} e^{ij_{\vp+\alpha \vp'}}\vert_{\alpha
= 0}$ one can verify that the limit $\ve \dar 0$ exists for the expectation
value with a total set of vectors and thus gives densely defined (not
closable) quadratic forms. They do not lead to operators but we know from (2.7)
that they define operator valued distributions for test functions from $H_1$.
Thus one could say that in the sense of operator valued
distributions the Urgleichung holds
$$
\frac{1}{i} \frac{\partial}{\partial x} \Psi(x) = \frac{\sqrt{2\pi\alpha}}{2}
\,[j(x),\Psi(x)]_+ . \eqno(4.2)
$$

The remarkable point is that the coupling constant $\lambda$ in
(1.1) is related to the statistics parameter $\alpha$. For fermions one
has a solution only for $\lambda = \sqrt{2\pi}$. Of course one could for
any $\lambda$ enforce fermi statistics by renormalizing the bare
fermion field $\psi \ra \sqrt{Z}\;\psi$, $j \ra Zj$ with a suitable
$Z(\lambda)$ but this just means pushing factors around. Alternatively
one could extend $\A_c$ by adding $e^{i\sqrt{2\pi\alpha}\,j_{\vp_\ve}}$, for all
$\alpha \in {\bf R_+}$. Then one gets in $\Ha_\omega$ uncountably many
orthogonal sectors, one for each $\alpha$, and in each sector a different
Urgleichung holds. Thus different anyons live in 
orthogonal Hilbert spaces and $e^{i\sqrt{2\pi\alpha}\,j_{\vp_\ve}}$ is not even 
weakly continuous in $\alpha$. If $\alpha$ is tied to $\lambda$ it is clear that
an expansion in $\lambda$ is doomed to failure and will never reveal
the true structure of the theory.
\vspace{0.6cm}

\section{Concluding remarks}
To summarize we gave a precise meaning to eq.(1.2a,b,c) by starting with bare
fermions, $\A = {\rm CAR}(\bf R)$. The shift $\tau_t$ is an automorphism of $\A$
which has KMS--states $\omega_\beta$ and associated representations
$\pi_\beta$. In $\pi_\beta(\A)''$ one finds bosonic modes $\A_c$ with an
algebraic structure independent on $\beta$. Taking the crossed product with an
outer automorphism of $\A_c$ or equivalently augmenting $\A_c$ by an unitary
operator to $\bar\A_c$ we discover in $\bar \pi_\beta(\A_c)''$ anyonic modes
which satisfy the Urgleichung in a distributional sense. For special values of
$\lambda$ they are dressed fermions distinct from the bare ones. From the
algebraic inclusions CAR({\it bare}) $\subset \pi_\beta(\A)'' \supset\A_c \subset
\bar \A_c \subset \bar\pi_\beta(\bar\A_c)'' \supset$ CAR({\it dressed}) one 
concludes
that in our model it cannot be decided whether fermions or bosons are more
fundamental. One can construct the dressed fermions either from bare fermions
or directly from the current algebra and our original question remains open
like the one whether the egg or the hen was first.
\vspace{0.6cm}

\section*{Acknowledgements}
The authors are grateful to H.Narnhofer for helpful discussions on the subject
of this paper.\\[2pt]

N.I. acknowledges the financial support from the ``Fonds zur F\"orderung der
wissenschaftlichen Forschung in \"Osterreich" under grant P11287--PHY and the
hospitality at the Institute for Theoretical Physics of University of Vienna.
\vspace{1cm}

\appendix
\newcounter{zahler}
\renewcommand{\thesection}{Appendix \Alph{zahler}:}
\renewcommand{\theequation}{\Alph{zahler}.\arabic{equation}}
\setcounter{zahler}{1}
\section{KMS--States --- Dirac sea and the Schwinger term}

An equilibrium state of a quantum system at finite temperature  $T = \beta^{-1}$
is characterized by the KMS--condition
\beq
\omega_\beta\left(\tau_t(A)\,B\right) = \omega_\beta(B\,\tau_{t+i\beta}A)
\eeq
with the time evolution $\tau_t$ as an automorphism of the algebra of
observables $\,\A\,$ analytically continued for imaginary times. Thus, an
equilibrium state for a system with an infinite number of free bosons can be
defined through the quasifree state over the algebra of smeared creation and
annihilation operators $a_f^\ast, a_g$,
$$
a_f^{(\ast)} =  \frac{1}{2\pi}\int_{-\infty}^{\infty} a^{(\ast)}(p)\tilde
f^{(\ast)}(p) dp
$$
so that for the non--smeared operators one has
\beq
\langle a^{\ast}(p)a(k)\rangle = 
\frac{2\pi\,p\, \delta(p\!-\!k)}{1 - e^{-\beta p}}
\eeq
similarly for fermions
\beq
\langle a^{\ast}(p)a(k)\rangle = 
\frac{2\pi\,\delta(p\!-\!k)}{1 + e^{\beta p}}
\eeq
Note that for (A.3) to be a well defined state there is no need for the
Hamiltonian to be bounded from below, in contrast to the $\,T=0\,$ case. There, a
Bogoliubov transformation is needed to ensure semiboundedness for the free
Hamiltonian. As has been realized already in the thirties \cite{J, BNN}, such a manipulation (corresponding to filling in the Dirac sea) leads to an
anomalous term in the current commutator --- eq.(2.13). One could be therefore
misleaded to think that the KMS--state ignores this anomaly. Actually, it
is the other way round --- the KMS--state automatically takes care for the
Dirac vacuum since for negative momenta (A.3) transforms into
$$
\langle a(p)a^{\ast}(k)\rangle = 
\frac{2\pi\,\delta(p\!-\!k)}{1 + e^{-\beta p}}
$$
and this corresponds exactly to exchanging the roles of creation and
annihilation operators.

Indeed, in momentum space, with
\beqa
\rho(p) = \int_{-\infty}^{\infty}\psi^\ast(x)\psi(x)e^{ipx} dx = 
\frac{1}{2\pi}\int_{-\infty}^{\infty}a^\ast(k+p) a(k) dk \no \\
\rho(-p) = \int_{-\infty}^{\infty}\psi^\ast(x)\psi(x)e^{-ipx} dx = 
\frac{1}{2\pi}\int_{-\infty}^{\infty}a^\ast(k)a(k+p) dk \no
\eeqa
$p\,$ always positive, one gets $(:\!\rho\!: = \rho - \langle \rho \rangle)$
\beqa
\langle :\!\rho(-p)\!: :\!\rho(p')\!:\rangle
= \int_{-\infty}^{\infty}\frac{dkdk'}{(2\pi)^2}
\langle a^\ast(k)a(k')\rangle \langle a(k+p)a^\ast(k'+p')\rangle = \no \\
= \left.\int_{-\infty}^{\infty} 
\frac{dk\,\delta(p\!-\!p')}{\left(1 + e^{\beta
k}\right)\left(1 + e^{-\beta(k+p)}\right)} = 
\frac{\delta(p\!-\!p')}{\beta \left(1 - e^{-\beta p}\right)}
\ln \frac{1 + e^{-\beta k}}{e^{\beta p} + e^{-\beta k}}
\right|_{-\infty}^{\infty} = \no 
\eeqa
\beq
= \frac{p}{1 - e^{-\beta p}}\,\delta(p\!-\!p') = 
F(p)\,\delta(p\!-\!p') 
\eeq
Then with the representation $\,\pi_\beta\,$ the following KMS--state over the
observables algebra $\A_c$ is accociated
$$
\omega_\beta(e^{ij_f}) = exp\left\{-\frac{1}{2}\int_{-\infty}^{\infty}
\frac{dp}{(2\pi)^2}\,\frac{p}{1-e^{-\beta p}}\,\vert\tilde f(p)\vert^2\right\}
$$
as follows from the general form of KMS--states over a Weyl algebra \cite{NT1}.

Similarly,
\beq
\langle :\!\rho(p')\!: :\!\rho(-p)\!:\rangle = 
-\frac{p}{1 - e^{\beta
p}}\,\delta(p\!-\!p') = F(-p)\,\delta(p\!-\!p')
\eeq
For $\,F(p)$ the following relation holds
\beq
F(-p) = e^{-\beta p}\,F(p) > 0 \quad \forall p \in \bf R
\eeq
With $\,\tau_t\rho(p) = e^{ipt}\rho(p) \longrightarrow e^{\beta p}\rho(p)\,$
and (A.6), validity of the KMS--condition, eq.(A.1), is verified
$$
\langle :\!\rho(-p)\!: :\!\,\tau_{i\beta}\rho(p')\!:\rangle = 
e^{-\beta p}F(p)\delta(p\!-\!p') = F(-p)\delta(p\!-\!p') = 
\langle :\!\rho(p')\!: :\!\rho(-p)\!:\rangle
$$
So, (A.4), (A.5) correspond to a KMS--state over a bosonic algebra and are both
temperature dependent. This is not the case for the commutator itself
$$
\langle[\rho(p), \rho(-p')]\rangle = F(p)\left(1 - e^{-\beta p}\right)
\delta(p\!-\!p') = p\,\delta(p\!-\!p')
$$
This is the well--known result from the $T=0$ case. Thus, the KMS--state for
$\,\beta > 0\,$ is by construction associated with the Dirac vacuum and the
current anomaly is recovered but it does not depend on the temperature (see
also \cite{HG}) despite the fact that the correlator functions do.
\vspace{0.6cm}
       
\addtocounter{zahler}{1}
\section{Non--commuting fields through crossed products}

The idea that the crossed product $C^\ast$--algebra extension is the tool that
makes possible construction of fermions (so, unobservable fields) from the
observable algebra has been first stated in \cite{DHR}. There, the problem of
obtaining different field groups has been shown to amount to construction of
extensions of the observable algebra by the group duals. Explicitly, crossed
products of $C^\ast$--algebras by semigroups of endomorphisms have been
introduced when proving the existence of a compact global gauge group in
particle physics given only the local observables \cite{DR}. Also in the
structural analysis of the symmetries in the algebraic QFT \cite{H}
extendibility of automorphisms from a unital $C^\ast$--algebra to its crossed
product by a compact group dual becomes of importance since it provides an
analysis of the symmetry breaking \cite{NT} and in the case of a broken
symmetry allows for concrete conlusions for the vacuum degeneracy \cite{BDLR}. 

The reason why a relatively complicated object --- crossed product over a
specially directed symmetric monoidal subcategory $\rm End \,\A$ of unital
endomorphisms of the observable algebra $\A$, is involved in considerations
in \cite{BDLR} is that in general, non--Abelian gauge groups are envisaged. For
the Abelian group $U(1)$ a significant simplification is possible since its
dual is also a group --- the group {\bf Z}. On the other hand, even in this
simple case the problem of describing the local gauge transformations remains
open. Therefore in the Abelian case consideration of crossed products over a
discrete group offers both a realistic framework and reasonable simplification
for the analysis of the resulting field algebra. We shall briefly outline the
general construction for this case, for more details see \cite{IN}.

We start with the CCR algebra $\A(\V_0,\sigma)$ over the real symplectic space
$\V_0$ with symplectic form $\sigma$, eq.(2.12), generated by the unitaries
$\,W(f), \,f \in \V_0$ with 
$$
W(f_1)W(f_2) = e^{i\sigma(f_1,f_2)}W(f_1+f_2),  
\qquad W(f)^\ast = W(-f) = W (f)^{-1}.
$$
Instead of the canonical extension $\,\bar\A(\V,\bar\sigma), \, \V_0\subset\V\,$
\cite{AMS}, we want to construct another algebra $\F$, such that ${\rm
CCR}(\V_0)\subset\F\subset{\rm CCR}(\V)$ and we choose $\,\V_0 = \C_0^{\,\infty},
\, \V = \partial^{-1}\C_0^{\,\infty}\,$. Any free (not inner) automorphism 
$\alpha, \, \alpha\in{\rm Aut}\,\A\,$ defines a crossed product $\, \F = \A\,
\st{\alpha}{\bowtie}\,{\bf Z}\,$. This may be thought as (see \cite{A}) adding to
the initial algebra $\A$ a single unitary operator $U$ together with all its
powers, so that one can formally write $ \,\F = \sum_n \A\,U^n\,$, with $U$
implementing the automorphism $\alpha$ in $\A, \,\, \alpha A = U\,A\,U^\ast,
\,\forall A\in\A$. Operator $\,U\,$ should be thought
of as a charge--creating operator and $\F$ is the minimal extension --- an
important point in comparison to the canonical extension which we find superfluous, 
especially when questions about statistical behaviour 
and time evolution are to be discussed. With the choice 
\beq
\alpha W(f) = e^{i\sigma(\bar g,f)}W(f), \qquad \bar g\in\V\backslash\V_0, 
\qquad \V_0\subset\V
\eeq
and identifying $U = W(\bar g)\,$, $\,\F\,$ is in a natural way a subalgebra 
of CCR$(\V)$.

If we take for $\A$ the current algebra $\A_c$ and for $U$ --- the idealized element
$U_\pi$ to be added to it, we find an obvious correspondence between the
functional picture from Sec.3 and the crossed product construction.
However, in the latter there is an additional structure present which makes it
in some cases favourable. Writing an element $F \in \F$ as $\, F = \sum_n A_n
U^n, \quad A_n \in \A\,$, we see that it is convenient to consider $\F$ as an
infinite vector space with $U^n$ as its basic unit vectors and $A_n =: (F)_n$
as components of $F$. The algebraic structure of $\F$ implies that
multiplication in this space is not componentwise but instead
$$
(F.G)_m = \dsum_n F_n\,\alpha^n\,G_{m-n}.
$$

Given a quasifree automorphism $\rho \in {\rm Aut}\,\A$, it can be extended 
to $\F$ iff the related automorphism $\gamma_\rho =
\rho\alpha\rho^{-1}\alpha^{-1}$ is inner for $\A$. Since $\gamma_\rho$ is
implemented by $W(\bar g_\rho - \bar g)$, this is nothing else but demanding
that $\bar g_\rho - \bar g \in \V_0$ and this is exactly the same requirement as
in the functional picture. This appears to be the case for the space 
translations and also
for the time evolution, but in the absence of long--range forces \cite{IN}.

Also a state $\omega(.)$
over $\A$ together with the representation $\,\pi_\omega\,$ associated with it 
through the GNS--construction can be extended to $\F$. The representation space
of $\F$ can be regarded as a direct sum of charge--$n$ subspaces, each of them
being associated with a state $ \omega\circ\alpha^{-n} $ and with $ \Ha_0 $, the
representation space of $ \A $, naturally imbedded in it. Since $\omega$ is
irreducible and $ \omega\circ\alpha^{-n} $ not normal with respect to it, the
extension of the state over $\A $ to a state over $\F$ is uniquely determined
by the expectation value with $\vert\Omega_0\rangle = \vert\omega\rangle$ in
this representation
$$
\langle\Omega_k\vert W^\ast(f) W(h) W(f)\vert\Omega_n\rangle = 
\delta_{kn} \, e^{-i\sigma (f+n\bar g, h)}\omega(W(h))
$$
where $ U_k\vert\Omega\rangle := \vert\Omega_k\rangle, \,
\langle\Omega_k\vert\Omega_n\rangle = \delta_{kn}\,$. This states nothing but
orthogonality of the different charge sectors, the same as in the functional
description, eq.(3.2).

In the crossed product gauge automorphism is naturally defined with
\beq
\gamma_\nu \,U^n = e^{2\pi i\nu n} U^n, \qquad \gamma_\nu\, W(f) = W(f)
\eeq
Thus for the representation $\pi_\Omega$ one finds
$$
\gamma_\nu\left(\vert F(f)^{(k)}\rangle\right) = 
\gamma_\nu\left(W(f)\vert\Omega_k\rangle\right) = 
e^{2\pi i\nu k}W(f)\vert\Omega_k\rangle,
$$
that justifies interpretation of the vectors  $\,\vert F(f)^{(k)}\rangle\,$
as belonging to the charge--$k$ subspace. However, $\A$ is a subalgebra of $\F$
for the gauge group $\T = [0,1)$, while it is a subalgebra of CAR for the gauge
group $\T\otimes\bf R$. Thus the crossed product algebra so constructed, being
really a Fermi algebra, does not coincide with CAR but is only contained in it.
In other words, such a type of extension does not allow incorporation also of
local gauge transformations which are of main importance in QFT.

Therefore we need a generalization of the construction in \cite{IN} which would
describe also the local gauge transformations. The most natural candidate for a
structural automorphism would be
\beq
\alpha_{\bar g_x} W(f) = e^{i\sum_{n=0}^{K}f^{(n)}(x)}W(f).
\eeq
However, it turns out that only for $K = 0$
the crossed product algebra so obtained allows for extension of space
translations as an automorphism of $\A$ --- the minimal requirement one should 
be able to meet. Already first derivative gives for the zero Fourier component 
of the difference $\bar g_{x_\delta} - \bar g_x$ an expression of the type
$\int y^{-1}\delta(y) dy$, so it drops out of $\,\C_0^{\,\infty}$. So,
the automorphism of interest reads
\beq
\alpha_{\bar g_x} W(f) = e^{if(x)}W(f)
\eeq
and can be interpreted as being implemented by $\,W(\bar g_x)$ with 
$\bar g_x = 2\pi\,\Theta (x\!-\!y)$.
Correspondingly, the operator we add to $\A$ through the crossed product is
\beq
U_x = e^{i 2\pi\int_{-\infty}^{x}j(y) dy}.
\eeq
Compared to \cite{IN} this means an enlargment of the test functions space
not with a kink but with its limit --- the sharp step function. In a
distributional sense it still can be considered as an element of  
$\,\partial^{-1}\V_0\,$ for some $\,\V_0\,$ since the derivative
of $\bar g_x$ has bounded zero Fourier component. Similarly, the extendibility
condition for space translations is found to be satisfied, $\bar
g_{x_\delta}-\bar g_x \in\V_0$ so that in the crossed product shifts are 
given by
\beq
\bar\tau_{x_\delta}U_x = V_{x_\delta}U_x, \qquad
V_{x_\delta} = W(\bar g_{x_\delta} - \bar g_x).
\eeq
Note that shifts do not commute with the structural automorphism $\alpha_{\bar
g_x}$, $\,\tau_{x_\delta}\alpha_{\bar g_x}W(f) \not= \alpha_{\bar
g_x}\tau_{x_\delta}W(f)$. Since 
\beq
\sigma(\bar g_x, \bar g_{x_\delta}) = -\pi\sgn(\delta),
\eeq
already the
elements of the first class are anticommuting and we identify $U_x =: \psi(x)$.
Then (B.4) (after smearing with a function from $\C_0^{\,\infty}$) is nothing
else but (1.5), i.e. the statement (or requirement) that currents generate
local gauge transformations of the so--constructed field. Any scaling of the
function which defines the structural automorphism $\,\alpha_{\bar g_x}\,$ 
destroys this
relation and fields obeying fractional statistics are obtained instead. This is
effectively the same as adding to the algebra $\,\A\,$ the element $U_\alpha$
with $\alpha = 2\pi\mu, \, \mu$ being the scaling parameter.

However, the crossed product offers one more interesting possibility: when for
the symplectic form in question instead of (B.7) (or its direct generalization
$\,\sigma(\bar g_x, \bar g_{x_\delta}) = (2n + 1)\pi, n \in \bf Z \,)$ another
relation takes place, $\,\sigma(\bar g_x, \bar g_{x_\delta}) = (2n+1)/\bar
n^2\,$ for some fixed $\,\bar n\in \bf Z$, the crossed product acquires a zone
structure, with $\,2n\bar n$--classes commuting, $\,(2n+1)$--classes 
anticommuting
and elements in the classes with numbers $m \in {\bf Z/Z}_{\bar n}$ obeying an
anyon statistics with parameter $\,r = \sqrt{2n + 1}\,m/\bar n\,$. So, fields 
with
different statistical behaviour are present in the same algebra, however the
Hilvbert space remains separable (which would not be the case if non--Abelian
group has been considered).
 
We want to emphasize that relation of the type $\,\psi(x+\delta_x) =
U_{x+\delta_x}\,$ may be misleading, the latter element 
exists in the crossed product only by eq.(B.6), so that for the derivative 
one finds
\beqa
\frac{\partial\psi(x)}{\partial x} := \lim_{\delta_x\ra 0}
\frac{\psi(x+\delta_x) - \psi(x)}{\delta_x} = 
\lim_{\delta_x\ra 0}\frac{1}{\delta_x}\left(V_{x_\delta} U_x - U_x\right) = \no
\\  
\lim_{\delta_x\ra 0}\frac{1}{\delta_x}\left(e^{i\,2\pi\,\delta_x\,j(x)} - 
1\right)U_x = 2\pi\,i\,j(x)U_x =: 2\pi\, i\,j(x) \psi(x).
\eeqa
This, together with (2.5) gives for the operators
\beq
i\psi_{f'} = \psi_{f}\, j_{\Theta'}.
\eeq
Note that in the crossed product, which can actually be considered as a left
$\A$--module, equations of motion (B.8), (B.9) appear (due to this reason) 
without an 
antisymmetrization, which was the case with the functional realization,
eq.(4.2), but otherwise the result is the same. Therefore the scaling 
sensitivity of the crossed product field algebra is another manifestation of
the quantum ``selection rule" for the value of $\,\lambda\,$ in the 
Urgleichung (1.2b).

\end{document}